\documentclass[12pt]{article}

\usepackage{amsmath,amssymb}
\usepackage[applemac]{inputenc}
\usepackage{amsmath,amssymb,fullpage}
\usepackage{showlabels}
\usepackage{color}

\newcommand{\dproof}{\noindent {Proof.} \quad}
\newcommand{\fproof}{\hfill $\square$ \bigskip}

\newtheorem{definition}{Definition}[section]
\newtheorem{example}{Example}[section]
\newtheorem{theorem}[definition]{Theorem}
\newtheorem{problem}[definition]{Problem}
\newtheorem{remark}[definition]{ \it Remark}

\newtheorem{proposition}[definition]{Proposition}
\newtheorem{lemma}[definition]{Lemma}

\numberwithin{equation}{section}

\def\1B{\text{1\!\!I}}

\begin{document}
\date{10 January 2018 }
\title{Viable Insider Markets}

\author{
Olfa Draouil$^{1,2}$ and Bernt \O ksendal$^{1,2}$}

\footnotetext[1]{Department of Mathematics, University of Oslo, P.O. Box 1053 Blindern, N--0316 Oslo, Norway.\\
Email: {\tt oksendal@math.uio.no}\\ \tt olfad@math.uio.no}

\footnotetext[2]{This research was carried out with support of the Norwegian Research Council, within the research project Challenges in Stochastic Control, Information and Applications (STOCONINF), project number 250768/F20.}

\maketitle



\begin{abstract} We consider the problem of optimal inside portfolio $\pi(t)$ in a financial market with a corresponding wealth process $X(t)=X^{\pi}(t)$ modelled by
\begin{align}\label{eq0.1}
  \begin{cases}
dX(t)&=\pi(t)X(t)[\alpha(t)dt+\beta(t)dB(t)]
; \quad t\in[0, T]\\
X(0)&=x_0>0,
\end{cases}
  \end{align}
where $B(\cdot)$ is a Brownian motion.
We assume that the insider at time $t$ has access to market information $\varepsilon_t>0$ units ahead of time, in addition to the history of the market up to time $t$. The problem is to find an insider portfolio $\pi^{*}$ which maximizes the expected logarithmic utility $J(\pi)$ of the terminal wealth, i.e. such that \\
$$\sup_{\pi}J(\pi)= J(\pi^{*}), \text {where } J(\pi)= \mathbb{E}[\log(X^{\pi}(T))].$$
The insider market is called \emph{viable} if this value is finite.\\
We study under what inside information flow $\mathbb{H}$ the insider market is viable or not.

 For example, assume that for all $t<T$ the insider knows the value of $B(t+\epsilon_t)$, where $t + \epsilon_t \geq T$ converges monotonically to $T$ from above as $t$ goes to $T$ from below.
Then (assuming that the insider has a perfect memory) at time $t$ she has the inside information $\mathcal{H}_t$,
consisting of the history $\mathcal{F}_t$ of $B(s); 0 \leq s \leq t$ plus all the values of Brownian motion in
the interval $[t+\epsilon_t, \epsilon_0]$, i.e. we have the enlarged filtration
\begin{equation}\label{eq0.2}
\mathbb{H}=\{\mathcal{H}_t\}_{t\in[0.T]},\quad \mathcal{H}_t=\mathcal{F}_t\vee\sigma(B(t+\epsilon_t+r),0\leq r \leq \epsilon_0-t-\epsilon_t), \forall t\in [0,T].
\end{equation}

 This gives rise to a progressive inside information flow $\mathbb{H}$. It is not clear if $B(\cdot)$ is a semimartingale under this filtration.
However, using forward integrals, Hida-Malliavin calculus and Donsker delta functionals we show that  if
$$\int_0^T\frac{1}{\varepsilon_t}dt=\infty,$$
then the insider market is not viable.
This extends a result in \cite{PK}, where it is proved that if the insider knows the value of $B(T)$ already from time 0 (corresponding to $\varepsilon_t=T-t$ for all $t$), then the market is not viable.
\end{abstract}

\section {Introduction}
The problem of insider trading is studied in many works for example \cite{A, AIS, AI, BO, C, DMOP2, DO1, H, IPW, PK}. This type of problems is related to the  enlargement of filtration \cite{Ja, J, Je, M}.\\
The purpose of this paper is to study the classical Merton problem of portfolio optimization in a financial market, extended to the case when the trader has inside information, i.e. has access to information about the value of random variables related to the terminal value $S(T)$ of the price of the risky asset.
\\

We are using basically the same framework as in the paper \cite{PK}, but with a progressive inside information flow rather than a fixed initial inside information.
We consider the inside information given by
\begin{equation}
\mathbb{H}=\{\mathcal{H}_t\}_{t\in[0.T]},\quad\mathcal{H}_t=\mathcal{F}_t\vee\sigma(B(t+\epsilon_t+r),0\leq r \leq \epsilon_0-t-\epsilon_t), \forall t\in [0,T],
\end{equation}
where $t+\epsilon_t>T$ and $t+\epsilon_t$ goes to $T$ when $t$ goes to $T$.\\
We show that if $t+\epsilon_t \geq T$ converges fast enough to $T$ as $t$ goes to $T,$ in the sense that
$$\int_0^T\frac{1}{\varepsilon_t}dt=\infty,$$
then the insider market is not viable.
This extends the result in \cite{PK}, where it is shown that if the trader knows the value of $B(T)$ from the beginning (corresponding to the case when $\epsilon_t = T-t$ for all $t$), then the value is infinite.\\

We conjecture that the converse is also true, namely that\\
the market is viable \emph{if and only if} $\int_0^T\frac{1}{\varepsilon_t}dt=\infty.$
In Section 3 we prove a result which supports this conjecture.
\\

We now describe this in more detail:\\
Consider a financial market where the unit price $S_0(t)$ of the risk free asset is
\begin{equation}\label{eq1.1}
    S_0(t)=1, \quad t\geq0
\end{equation}
and the unit price process $S(t)$ of the risky asset has no jumps and is given by

\begin{align}\label{eq1.2}
\begin{cases}
 dS(t) &= S(t) [\alpha(t,Y_t) dt + \beta(t,Y_t) dB(t)]
 ; \quad t\geq0\\
    S(0)&>0.
\end{cases}
\end{align}
Here $B(t)$ is a Brownian motion on a given filtered probability space $(\Omega, \mathbb{F}=\{\mathcal{F}_t\}_{0\leq t\leq T},P)$, where $\mathcal{F}_t$ is the $\sigma$-algebra generated by $B(s); 0 \leq s \leq t$.
For each $t \in [0,T]$ we fix a constant $\varepsilon_t >0 $ and we let $Y_t$ be a given $\sigma(B(t+\epsilon_t))$-measurable random variable representing the inside information available to the controller. We assume that the insider has a perfect memory and thus she at time $t$ knows $\mathcal{F}_t$ and can remember all the previous values of $Y_s; 0 \leq s \leq t$. Accordingly, we consider the enlarged filtration $\mathbb{G}=\{\mathcal{G}_t\}_{t \geq 0}$ given by $\mathcal{G}_t=\mathcal{F}_t \vee \sigma(Y_{s}; 0 \leq s \leq t)$. In addition, we assume that for each $t$ the random variable $Y_t$ has a Donsker delta functional $\delta_{Y_t}(y)$.

It is not clear if $B(t)$ is a semimartingale with respect to $\mathbb{G}$. Therefore we interpret the $dB-$ integral in equation \eqref{eq1.2} as a \emph{forward integral}. See Section 2.\\

Now suppose we apply a self-financing  $\mathcal{G}_t$-adapted portfolio $\pi(t)=\pi(t,Y_t)$ to this market, where $\pi(t)$ is the fraction of the corresponding wealth at time $t$ invested in the risky asset. Then the corresponding wealth process $X(t)=X^{\pi}(t)$ satisfies the equation
\begin{align}\label{eq1.3}
  \begin{cases}
dX(t)&=\pi(t,Y_t)X(t)[\alpha(t)dt+\beta(t)dB(t)
; \quad t\in[0, T]\\
X(0)&=x_0>0.
\end{cases}
  \end{align}

Let $\mathcal{A}$ be the set of $\mathbb{G}-$adapted, self-financing portfolios $\pi(t,Y_t)$ such that the forward equation \eqref{eq1.3} has a unique solution. We study the following insider optimal portfolio problem:

\begin{problem}
Find $\pi^{\ast}\in \mathcal{A}$ such that
\begin{equation}\label{eq17}
J(\pi^*) =\sup_{\pi\in\mathcal{A}} J(\pi),
\end{equation}
where
\begin{equation}\label{eq18}
J(\pi):=  \mathbb{E}[\log(X^\pi(T))].
\end{equation}
\end{problem}
\begin{definition}
The market \eqref{eq1.1}-\eqref{eq1.3} is called \emph{viable} if
\begin{equation}\label{eq via}
\sup_{\pi\in\mathcal{A}} J(\pi) < \infty.
\end{equation}
\end{definition}

Later in this paper we will treat the special case when $Y_t=B(t+\epsilon_t).$ Then instead of the general filtration $\mathbb{G}$ we will consider the filtration $\mathbb{H}=\{\mathcal{H}_t\}_{t\in[0.T]}$, where $\mathcal{H}_t=\mathcal{F}_t\vee\sigma(B(t+\epsilon_t+r),0\leq r \leq \epsilon_0-t-\epsilon_t) \forall t\in [0,T]$,  to deal with the problem \eqref{eq1.1}-\eqref{eq via}.
Note that in this case we have that $\mathcal{A}$ is the set of $\mathbb{H}$-adapted, self-financing portfolios $\pi(t,B(t+\epsilon_t))$ such that the equation \eqref{eq1.3} with $Y_t=B(t+\epsilon_t)$ has a unique solution.

The purpose of this paper is to study under what inside information flow $\mathbb{H}$ the insider market is viable or not. In particular, we show that if for all $t$  the insider knows the value of $B(t+\varepsilon_t)$ for some $\varepsilon_t>0$, then the insider market is not viable if
$$\int_0^T\frac{1}{\varepsilon_t}dt=\infty.$$
This extends a result in \cite{PK}, where it is proved that if the insider knows the value of $B(T)$ already from time 0 (corresponding to the case $\varepsilon_t=T-t$ for all $t$), then the market is not viable.\\

This type of problem has been studied in the paper \cite{DOPP} (see in particular Section 6.3), but the methods there are different and the results are no as explicit as in our paper.\\
Also Corcuera et al. \cite{C} treated this type of problem but using another kind of filtration called dynamical enlargement of filtration. This filtration consists in the knowledge of a functional of the underlying deformed by an independent noise process which tends to 0 as $T$ approaches.
This type of information is different from the one that we consider in this paper. The technique also of treating the problem is different from \cite{C}.\\
This paper is  organized as follows:

In Section 2, we give a brief review of forward integrals, Donsker delta functional and Hida-Malliavin calculus.

In  Section 3, we suppose that $Y_t=(Y_t^{(1)},Y_t^{(2)})$.
In Theorem 3.1 we give an explicit expression of the optimal portfolio in terms of the conditional expectation of the Donsker of $Y_t$ and its Hida-Malliavin derivative.
Then we consider the case where $Y_t^{(1)}=B(t+\varepsilon_t^{(1)})$ and $Y_t^{(2)}=B(t+\varepsilon_t^{(2)})$ for $\varepsilon_t^{(1)}<\varepsilon_t^{(2)}$. In Theorem 3.2 we prove that the maximal expected logarithmic utility of the terminal wealth process depends only on the knowledge of $B(t+\varepsilon_t^{(1)})$. Therefore we deduce that given the information $\sigma(B(t+\epsilon_t+r),0\leq r \leq \epsilon_0-t-\epsilon_t), \forall t\in [0,T]$ for the insider, the  information of $B(t+\epsilon_t)$ is relevant.
In Theorem 3.4 we give  a  condition in terms of $\epsilon_t $ in order to obtain a viable market.
Then we give some examples of characterization of viability.

Finally, in Subsection 3.2 we treat the case where $t+\epsilon_t<T$ and $\mathcal{H}_t=\mathcal{F}_{t+\epsilon_t}$. We show that this insider market is viable if and only if $\int_0^T\frac{1}{\epsilon_t}dt<\infty$. Hence the market is never viable in this case.

\section {Forward integrals and the Donsker delta functional}

\subsection{ The forward integral with respect to Brownian motion}
Since we are not sure that $B(t)$ is a semimartingale under $\mathbb{H}$ or not we interpret the equations \eqref{eq1.2} and \eqref{eq1.3} as forward integrals.
The forward integral with respect to Brownian motion was first defined in
the seminal paper \cite{RV} and further studied in \cite{RV1}, \cite{RV2}. This integral was
introduced in the modelling of insider trading in \cite{BO} and then applied by several authors
in questions related to insider trading and stochastic control with
advanced information (see, e.g., \cite{DMOP2}). The forward integral was later extended to Poisson random measure integrals in \cite{DMOP1}.

\begin{definition}
We say that a stochastic process $\phi = \phi(t), t\in[0, T ]$, is
\emph{forward integrable} (in the weak sense) over the interval $[0, T ]$ with respect to
$B$ if there exists a process $I = I(t), t\in[0, T ]$, such that
\begin{equation}
\sup_{t\in[0,T ]}|\int_0^t\phi(s)\frac{B(s+\epsilon)-B(s)}{\epsilon}ds-I(t)|\rightarrow 0, \quad \epsilon\rightarrow0^+,
\end{equation}
in probability. In this case we write
\begin{equation}
I(t) :=\int_0^t\phi(s)d^-B(s), t\in[0, T ],
\end{equation}
and call $I(t)$ the \emph{forward integral} of $\phi$ with respect to $B$ on $[0, t]$.
\end{definition}

The following results give a more intuitive interpretation of the forward integral
as a limit of Riemann sums:
\begin{lemma}
Suppose $\phi$ is c\`{a}gl\`{a}d and forward integrable. Then
\begin{equation}
\int_0^T\phi(s)d^-B(s) = \lim _{\triangle t\rightarrow0}\sum_{j=1}^{J_n}\phi(t_{j-1})(B(t_j)-B(t_{j-1})),
\end{equation}
with convergence in probability. Here the limit is taken over the partitions \\
$0 =t_0 < t_1 < ... < t_{J_n}= T$ of $t\in[0, T ]$ with $\triangle t:= \max_{j=1,...,J_n}(t_j- t_{j-1})\rightarrow 0,
n\rightarrow\infty.$
\end{lemma}

\begin{remark}

From the previous lemma we can see that, if the integrand $\phi$
is $\mathcal{F}$-adapted, then the Riemann sums are also an approximation to the It\^{o}
integral of $\phi$ with respect to the Brownian motion. Hence in this case the
forward integral and the It\^{o} integral coincide. In this sense we can regard
the forward integral as an extension of the It\^{o} integral to a nonanticipating
setting.
\end{remark}

We now give some useful properties of the forward integral. The
following result is an immediate consequence of the definition.

\begin{lemma}
 Suppose $\phi$ is a forward integrable stochastic process and $G$ a
random variable. Then the product $G\phi$ is forward integrable stochastic process and
\begin{equation}
\int_0^TG\phi(t)d^-B(t) = G\int_0^T\phi(t)d^-B(t).
\end{equation}
\end{lemma}

The next result shows that the forward integral is an extension of the
integral with respect to a semimartingale:
\begin{lemma}
Let $\mathbb{G} := \{\mathcal{G}_t, t\in[0, T ]\} (T > 0)$ be a given filtration. Suppose
that
\begin{enumerate}
\item $B$ is a semimartingale with respect to the filtration $\mathbb{G}$.
\item $\phi$ is $\mathbb{G}$-predictable and the integral
\begin{equation}
\int_0^T\phi(t)dB(t),
\end{equation}
with respect to $B$, exists.\\
Then $\phi$ is forward integrable and
\begin{equation}
\int_0^T\phi(t)d^-B(t)=\int_0^T\phi(t)dB(t),
\end{equation}
\end{enumerate}
\end{lemma}

We now turn to the It\^{o} formula for forward integrals. In this connection it is
convenient to introduce a notation that is analogous to the classical notation
for It\^{o} processes.
\begin{definition}
A \emph{forward process} (with respect to $B$) is a stochastic process
of the form
\begin{equation}\label{forward form 1}
X(t) = x +\int_0^tu(s)ds +\int_0^tv(s)d^-B(s), \quad t\in[0, T ],
\end{equation}
($x$ constant), where $\int_0^T|u(s)|ds <\infty, \mathbf{P}$-a.s.
and $v$ is a forward integrable stochastic process. A shorthand notation for
(\ref{forward form 1}) is that
\begin{equation}
d^-X(t) = u(t)dt + v(t)d^-B(t).
\end{equation}
\end{definition}

\begin{theorem}{\emph{The one-dimensional It\^{o} formula for forward integrals.}} \\
Let
\begin{equation}
d^-X(t) = u(t)dt + v(t)d^-B(t),
\end{equation}
be a forward process. Let $f\in\mathbf{C}^{1,2}([0, T ]\times\mathbb{R})$ and define
\begin{equation}
Y (t) = f(t,X(t)), \quad t\in[0, T ].
\end{equation}
Then $Y (t), t\in[0, T ]$, is also a forward process and
\begin{equation}
d^-Y (t) = \frac{\partial f}{\partial t}(t,X(t))dt+\frac{\partial f}{\partial x}(t,X(t))d^-X(t)+\frac{1}{2}\frac{\partial^2f}{\partial x^2} (t,X(t))v^2(t)dt.
\end{equation}
\end{theorem}

We also need the following forward integral result, which is obtained by an adaptation of the proof of Theorem 8.18 in \cite{DOP}:
\begin{proposition}

Let $\varphi$ be a c\`{a}gl\`{a}d and forward integrable process in $L^2(\lambda \times P)$. Then
$$\mathbb{E}[D_{s^{+}} \varphi(s)|\mathcal{F}_s]:= \lim _{\epsilon \rightarrow 0^{+}} \frac{1}{\epsilon}\int_{s-\epsilon}^s \mathbb{E}[D_s \varphi(t)|\mathcal{F}_s]dt,$$
exists in $L^2(\lambda \times P)$ and
\begin{align}\label{eq2.12}
& \mathbb{E}[\int_0^T \varphi(s) d^{-}B(s)] = \mathbb{E}[\int_0^T \mathbb{E}[D_{s^{+}} \varphi(s)|\mathcal{F}_s] ds].
\end{align}
where $D_s\varphi$ is the Hida-Malliavin derivative of $\varphi$ with respect to the Brownian motion $B$.
\end{proposition}

Similar definitions and results can be obtained in the Poisson random measure case. See \cite{DMOP1} and \cite{DOP}.

\subsection{The Donsker delta functional}
In this section we will define the Donsker delta function of a random variable $Y=(Y^1,Y^2)$ with value in $\mathbb{R}^2$,
because in section 3 we want to prove that under the filtration $\mathbb{H}$ the knowledge of the value of $B(t+\epsilon_t)$ is required to study the viability of the market. That is mean we will prove that if we know the values of  $B(t+\epsilon^{(1)}_t)$ and $B(t+\epsilon^{(2)}_t)$ for $\epsilon_t^{(1)}<\epsilon^{(2)}_t$ then the maximal expected utility of the terminal wealth depends only on the knowledge of $B(t+\epsilon^{(1)}_t)$.
\begin{definition}
Let $Y=(Y^1,Y^2) :\Omega\rightarrow\mathbb{R}^2$ be a random variable which also belongs to the Hida space $(\mathcal{S})^{\ast}$ of stochastic distributions. Then a continuous functional
\begin{equation}\label{donsker}
    \delta_Y(.): \mathbb{R}^2\rightarrow (\mathcal{S})^{\ast}
\end{equation}
is called a \emph{Donsker delta functional} of $Y $ if it has the property that
\begin{equation}\label{donsker property }
    \int_{\mathbb{R}^2}g(y)\delta_Y(y)dy = g(Y) \quad a.s.
\end{equation}
for all (measurable) $g : \mathbb{R}^2 \rightarrow \mathbb{R}$ such that the integral converges. Here, and in the
following, $dy = dy_1 dy_2$ denotes 2-dimensional Lebesgue measure.
\end{definition}
\begin{proposition}\cite{AaOU}
Suppose $Y : \Omega\rightarrow\mathbb{R}^2 $ is a normally distributed random variable
with mean $m = \mathbb{E}[Y]$ and covariance matrix $C = (c_{ij})_{1\leq i,j\leq 2}$. Suppose $C$ is
invertible with inverse $A =(a_{ij})_{1\leq i,j\leq 2}$ . Then $\delta_Y (y)$ is unique and is given by the
expression
\begin{equation}
\delta_Y (y) = (2\pi)^{-1}|A| \exp^{\diamond}\{-\frac{1}{2}\sum_{i,j=1}^2a_{ij} (y_i-Y_i)\diamond(y_j-Y_j )\},
\end{equation}
where $|A|$ is the determinant of $A$
\end{proposition}





Explicit formulas for the Donsker delta functional are known in many cases. For the Gaussian case, see Section 2.3. For details and more general cases, see e.g. \cite{AaOU}, \cite{DiO1},\cite{DiO2},\cite{MOP} and \cite{DO1}.

In the next subsection we choose the particular example $Y(t)= (B(t+\epsilon_t^{(1)}),B(t+\epsilon_t^{(2)}))$  where $\epsilon_t^{(1)} <\epsilon_t^{(2)}$ because in Section 3 we aim to prove that the knowledge of $B(t+\epsilon_t^{(1)})$ is required to study the viability of the market.

Here are some useful formulas used in the calculation of Example 2.1:\\
Let $F$ and $G$ $\in(\mathcal{S})^{\ast}$, we have:

\begin{equation}
D_t(F\diamond G)=F\diamond D_tG+D_tF\diamond G,
\end{equation}
\begin{equation}
D_t(F^{\diamond n})=nF^{\diamond (n-1)}\diamond D_t F, \quad (n=1,2...),
\end{equation}
\begin{equation}
D_t\exp^{\diamond}F=\exp^{\diamond}F\diamond D_t F,
\end{equation}
\begin{equation}
\mathbb{E}[\exp^{\diamond}F| \mathcal{F}_t]=\exp{\diamond}\mathbb{E}[F| \mathcal{F}_t],
\end{equation}
\begin{equation}
D_t(\int_{\mathbb{R}}f(s)dB(s))=f(t).
\end{equation}

\subsection{Examples}
\begin{example}
Let $Y(t)= (B(t+\epsilon_t^{(1)}),B(t+\epsilon_t^{(2)}))$  where $\epsilon_t^{(1)} <\epsilon_t^{(2)}$. The expectation of $Y$ is given by $\mathbb{E}[Y(t)]=(0,0)$ and the variance matrix is
 \[
   V=
  \left[ {\begin{array}{cc}
   t+\epsilon_t^{(1)} & t+\epsilon_t^{(1)} \\
   t+\epsilon_t^{(1)} & t+\epsilon_t^{(2)} \\
  \end{array} } \right].
\]
Its inverse matrix is given by
\[
   A=
  \left[ {\begin{array}{cc}
   \frac{t+\epsilon_t^{(2)}}{(t+\epsilon_t^{(1)})(\epsilon_t^{(2)}-\epsilon_t^{(1)})} & \frac{-1}{\epsilon_t^{(2)}-\epsilon_t^{(1)}}\\
   \frac{-1}{\epsilon_t^{(2)}-\epsilon_t^{(1)}} & \frac{1}{\epsilon_t^{(2)}-\epsilon_t^{(1)}} \\
  \end{array} } \right].
\]
The determinant of A is given by:
\begin{equation}
det(A)=\frac{1}{(t+\epsilon_t^{(1)})(\epsilon_t^{(2)}-\epsilon_t^{(1)})}>0.
\end{equation}
Then the Donsker delta of $Y(t)= (B(t+\epsilon_t^{(1)}),B(t+\epsilon_t^{(2)}))$ is given by:
\begin{align}
&\delta_{B(t+\epsilon_t^{(1)}),B(t+\epsilon_t^{(2)})}=(2\pi)^{-1}\sqrt{det(A)}
\exp^{\diamond}\Big(-\frac{t+\epsilon_t^{(2)}}{2(t+\epsilon_t^{(1)})(\epsilon_t^{(2)}-\epsilon_t^{(1)})}(y_1-B(t+\epsilon_t^{(1)}))^{\diamond 2}\nonumber\\
&+\frac{1}{\epsilon_t^{(2)}-\epsilon_t^{(1)}}(y_1-B(t+\epsilon_t^{(1)}))\diamond (y_2-B(t+\epsilon_t^{(2)}))
-\frac{1}{2(\epsilon_t^{(2)}-\epsilon_t^{(1)})}(y_2-B(t+\epsilon_t^{(2)}))^{\diamond 2}]\Big).
\end{align}
Using the Wick rule when taking conditional expectation and using the martingale property of the process $B(t+\epsilon_t^{(1)})$ and $B(t+\epsilon_t^{(2)})$ we get
\begin{align}
&\mathbb{E}[\delta_Y (y)|\mathcal{F}_t]=(2\pi)^{-1}\sqrt{det(A)}
\exp^{\diamond}(\mathbb{E}[-\frac{t+\epsilon_t^{(2)}}{2(t+\epsilon_t^{(1)})(\epsilon_t^{(2)}-\epsilon_t^{(1)})}
(y_1-B(t+\epsilon_t^{(1)}))^{\diamond 2}|\mathcal{F}_t]\nonumber\\
&+\mathbb{E}[\frac{1}{\epsilon_t^{(2)}-\epsilon_t^{(1)}}(y_1-B(t+\epsilon_t^{(1)}))\diamond (y_2-B(t+\epsilon_t^{(2)}))|\mathcal{F}_t]\nonumber\\
&-\mathbb{E}[\frac{1}{2(\epsilon_t^{(2)}-\epsilon_t^{(1)})}(y_2-B(t+\epsilon_t^{(2)}))^{\diamond 2}|\mathcal{F}_t])\nonumber\\
&=(2\pi)^{-1}\sqrt{det(A)}
\exp^{\diamond}(-\frac{t+\epsilon_t^{(2)}}{2(t+\epsilon_t^{(1)})(\epsilon_t^{(2)}-\epsilon_t^{(1)})}
(y_1-B(t))^{\diamond 2}\nonumber\\
&+\frac{1}{\epsilon_t^{(2)}-\epsilon_t^{(1)}}(y_1-B(t))\diamond (y_2-B(t))-\frac{1}{2(\epsilon_t^{(2)}-\epsilon_t^{(1)})}(y_2-B(t))^{\diamond 2})\nonumber\\
&=(2\pi)^{-1}\sqrt{det(A)}\exp(-\frac{(y_1-y_2)^2}{2(\epsilon_t^{(2)}-\epsilon_t^{(1)})})\exp^{\diamond}(-\frac{1}{2(t+\epsilon_t^{(1)})}(y_1-B(t))^{\diamond 2}).
\end{align}
To be with same notation as in \cite{AaOU},
we denote by  $a=-\frac{1}{2 (t+\epsilon_t^{(1)})}$ and $\psi=1_{[0,t]}$.\\
We have $2|a|\|\psi\|=\frac{t}{t+\epsilon_t^{(1)}}<1$ then using Corollary 3.6 in \cite{AaOU} we get
\begin{align}
&\mathbb{E}[\delta_Y (y)|\mathcal{F}_t]=(2\pi)^{-1}\exp(-\frac{(y_1-y_2)^2}{2(\epsilon_t^{(2)}-\epsilon_t^{(1)})})
\sqrt{\frac{1}{\epsilon_t^{(1)}(\epsilon_t^{(2)}-\epsilon_t^{(1)})}}\nonumber\\
&\exp(-\frac{1}{2\epsilon_t^{(1)}}(y_1-B(t))^2).
\end{align}
Now we want to compute  $\mathbb{E}[D_t\delta_Y (y)|\mathcal{F}_t]$.
We have
\begin{align}
&D_t\delta_Y (y)=D_t[(2\pi)^{-1}\sqrt{det(A)}
\exp^{\diamond}\Big(-\frac{t+\epsilon_t^{(2)}}{2(t+\epsilon_t^{(1)})(\epsilon_t^{(2)}-\epsilon_t^{(1)})}(y_1-B(t+\epsilon_t^{(1)}))^{\diamond 2}\nonumber\\
&+\frac{1}{\epsilon_t^{(2)}-\epsilon_t^{(1)}}(y_1-B(t+\epsilon_t^{(1)}))\diamond (y_2-B(t+\epsilon_t^{(2)}))
-\frac{1}{2(\epsilon_t^{(2)}-\epsilon_t^{(1)})}(y_2-B(t+\epsilon_t^{(2)}))^{\diamond 2}]\Big)]\nonumber\\
&=(2\pi)^{-1}\sqrt{det(A)}\exp^{\diamond}\Big(-\frac{t+\epsilon_t^{(2)}}{2(t+\epsilon_t^{(1)})(\epsilon_t^{(2)}-\epsilon_t^{(1)})}(y_1-B(t+\epsilon_t^{(1)}))^{\diamond 2}\\
&+\frac{1}{\epsilon_t^{(2)}-\epsilon_t^{(1)}}(y_1-B(t+\epsilon_t^{(1)}))\diamond (y_2-B(t+\epsilon_t^{(2)}))
-\frac{1}{2(\epsilon_t^{(2)}-\epsilon_t^{(1)})}(y_2-B(t+\epsilon_t^{(2)}))^{\diamond 2}]\Big)\\
&\diamond\{\frac{t+\epsilon_t^{(2)}}{(t+\epsilon_t^{(1)})(\epsilon_t^{(2)}-\epsilon_t^{(1)})}(y_1-B(t+\epsilon_t^{(1)}))\\
&-\frac{1}{\epsilon_t^{(2)}-\epsilon_t^{(1)}}(y_1-B(t+\epsilon_t^{(1)})+y_2-B(t+\epsilon_t^{(2)}))+\frac{1}{\epsilon_t^{(2)}-\epsilon_t^{(1)}}(y_2-B(t+\epsilon_t^{(2)}))\}.
\end{align}
Then
\begin{align}
&\mathbb{E}[D_t\delta_Y (y)|\mathcal{F}_t]=(2\pi)^{-1}\sqrt{det(A)}\mathbb{E}[\exp^{\diamond}\Big(-\frac{t+\epsilon_t^{(2)}}{2(t+\epsilon_t^{(1)})(\epsilon_t^{(2)}-\epsilon_t^{(1)})}(y_1-B(t+\epsilon_t^{(1)}))^{\diamond 2}\nonumber\\
&+\frac{1}{\epsilon_t^{(2)}-\epsilon_t^{(1)}}(y_1-B(t+\epsilon_t^{(1)}))\diamond (y_2-B(t+\epsilon_t^{(2)}))
-\frac{1}{2(\epsilon_t^{(2)}-\epsilon_t^{(1)})}(y_2-B(t+\epsilon_t^{(2)}))^{\diamond 2}]\Big)|\mathcal{F}_t]\\
&\diamond \mathbb{E}[\frac{t+\epsilon_t^{(2)}}{(t+\epsilon_t^{(1)})(\epsilon_t^{(2)}-\epsilon_t^{(1)})}(y_1-B(t+\epsilon_t^{(1)}))\\
&-\frac{1}{\epsilon_t^{(2)}-\epsilon_t^{(1)}}(y_1-B(t+\epsilon_t^{(1)})+y_2-B(t+\epsilon_t^{(2)}))+\frac{1}{\epsilon_t^{(2)}-\epsilon_t^{(1)}}(y_2-B(t+\epsilon_t^{(2)}))|\mathcal{F}_t]\nonumber\\
&=(2\pi)^{-1}\sqrt{\frac{1}{(t+\epsilon_t^{(1)})(\epsilon_t^{(2)}-\epsilon_t^{(1)})}}\exp(-\frac{(y_1-y_2)^2}{2(\epsilon_t^{(2)}-\epsilon_t^{(1)})})\exp^{\diamond}(-\frac{-(\epsilon_t^{(2)}-\epsilon_t^{(1)})}{2(\epsilon_t^{(2)}-\epsilon_t^{(1)})(t+\epsilon_t^{(1)})}(y_1-B(t))^{\diamond 2})\nonumber\\
&\diamond \{\frac{y_1-B(t)}{t+\epsilon_t^{(1)}} \}.
\end{align}
By Lemma 3.8 in \cite{AaOU}
\begin{align}
&\frac{1}{\sqrt{t+\epsilon_t^{(1)}}}\exp^{\diamond}(-\frac{1}{2(t+\epsilon_t^{(1)})}(y_1-B(t))^{\diamond 2})\diamond\frac{y_1-B(t)}{t+\epsilon_t^{(1)}}\\
&=\frac{1}{\sqrt{\epsilon_t^{(1)}}}\exp(-\frac{(y_1-B(t))^{2}}{2\epsilon_t^{(1)}})\frac{y_1-B(t)}{\epsilon_t^{(1)}}.
\end{align}
Then
\begin{align}
&\mathbb{E}[D_t\delta_Y (y)|\mathcal{F}_t]=(2\pi)^{-1}\exp(-\frac{(y_1-y_2)^2}{2(\epsilon_t^{(2)}-\epsilon_t^{(1)})})
\sqrt{\frac{1}{\epsilon_t^{(1)}(\epsilon_t^{(2)}-\epsilon_t^{(1)})}}\nonumber\\
&\exp(-\frac{(y_1-B(t))^2}{2\epsilon_t^{(1)}})\frac{y_1-B(t)}{\epsilon_t^{(1)}}.
\end{align}
\end{example}

\section {The main results}
We first recall how to find a general expression for the optimal portfolio with $Y_t=(Y^1_t,Y^2_t)$.
This was first proved in \cite{DO1}. See also \cite{OR}.
We now return to Problem 1.1. We normalize $x_0$ to be 1, so that we have
\begin{equation}
\begin{cases}
dX(t)=\pi(t,Y_t)X(t)[\alpha(t) dt+\beta(t) dB(t)]\nonumber\\
X(0)=1
\end{cases}
\end{equation}
where $Y_t=(Y^1_t,Y^2_t)$.

The solution of this stochastic differential equation is given by:
\begin{equation}
X(t)=\exp[\int_0^t\pi(s,Y_s)\beta(s) dB(s)+\int_0^t(\pi(s,Y_s)\alpha(s)-\frac{1}{2}\pi^2(s,Y_s)\beta^2(s))ds].
\end{equation}
Hence the performance functional \eqref{eq18} is given by:
\begin{align}\label{eq0.2}
J(\pi)&=\mathbb{E}[\log X^{\pi}(T)]\nonumber\\
&=\mathbb{E}[\int_0^T\pi(t,Y_t)\beta(t) dB(t)+\int_0^T(\pi(t,Y_t)\alpha(t)-\frac{1}{2}\pi^2(t,Y_t)\beta^2(t))dt].
\end{align}

We now use the definition of the Donsker delta functional of $\delta_{Y_t}(y)$ we get that:
\begin{align}
&\int_0^T\pi(t,Y_t)\beta(t) dB(t)+\int_0^T(\pi(t,Y_t)\alpha(t)-\frac{1}{2}\pi^2(t,Y_t)\beta^2(t))dt\nonumber\\
&=\int_{\mathbb{R}^2}\int_0^T\pi(t,y)\beta(t)\delta_{Y_t}(y) dB(t)dy+\int_{\mathbb{R}^2}\int_0^T(\pi(t,y)\alpha(t)-\frac{1}{2}\pi^2(t,y)\beta^2(t))\delta_{Y_t}(y)dtdy,
\end{align}
then using the following general duality formula for forward integrals:
\begin{equation}
\mathbb{E}[\int_0^T \phi(t)dB(t)]=\mathbb{E}[\int_0^T\mathbb{E}[D_{t^+}\phi(t)|\mathcal{F}_t] dt],
\end{equation}
then we get:
\begin{align}\label{eq0.5}
&\mathbb{E}[\log X^{\pi}(T)]
=\mathbb{E}[\int_{\mathbb{R}^2}(\int_0^T\pi(t,y)\beta(t)\mathbb{E}[D_{t^+}\delta_{Y_t}(y)|\mathcal{F}_t] dt)dy]\nonumber\\
&+\mathbb{E}[\int_{\mathbb{R}^2}(\int_0^T(\pi(t,y)\alpha(t)-\frac{1}{2}\pi^2(t,y)\beta^2(t))\mathbb{E}[\delta_{Y_t}(y)|\mathcal{F}_t]dt)dy]\nonumber\\
&=\int_{\mathbb{R}^2}\mathbb{E}[\int_0^T\{\pi(t,y)\beta(t)\mathbb{E}[D_{t^+}\delta_{Y_t}(y)|\mathcal{F}_t]+\pi(t,y)\alpha(t) \mathbb{E}[\delta_{Y_t}(y)|\mathcal{F}_t]\nonumber\\
&-\frac{1}{2}\pi^2(t,y)\beta^2(t)\mathbb{E}[\delta_{Y_t}(y)|\mathcal{F}_t]\}dt]dy.
\end{align}

The map
\begin{equation}
\pi\mapsto \pi \beta(t)\mathbb{E}[D_{t^+}\delta_{Y_t}(y)|\mathcal{F}_t]+\pi\alpha(t)\mathbb{E}[\delta_{Y_t}(y)|\mathcal{F}_t]
-\frac{1}{2}\pi^2\beta^2(t)\mathbb{E}[\delta_{Y_t}(y)|\mathcal{F}_t]
\end{equation}
is maximal when
\begin{equation}
\beta(t)\mathbb{E}[D_{t^+}\delta_{Y_t}(y)|\mathcal{F}_t]+\alpha(t)\mathbb{E}[\delta_{Y_t}(y)|\mathcal{F}_t]
-\hat{\pi}(t,y)\beta^2(t)\mathbb{E}[\delta_{Y_t}(y)|\mathcal{F}_t]=0,
\end{equation}
i.e. when
\begin{equation}\label{eq0.8}
\pi=\hat{\pi}(t,y)=\frac{\alpha(t)}{\beta^2(t)}+\frac{\mathbb{E}[D_{t^+}\delta_{Y_t}(y)|\mathcal{F}_t]}{\beta(t)\mathbb{E}[\delta_{Y_t}(y)|\mathcal{F}_t]}.
\end{equation}
We summarize what we have proved as follows:
\begin{theorem}\cite{DO1}, \cite{OR}
The optimal insider portfolio $\pi^*$ of Problem 1.1 is given by
\begin{equation}
\pi^*(t,Y_t)=\frac{\alpha(t)}{\beta^2(t)}+\frac{\mathbb{E}[D_{t^+}\delta_{Y_t}(y)|\mathcal{F}_t]_{y=Y_t}}{\beta(t)\mathbb{E}[\delta_{Y_t}(y)|\mathcal{F}_t]_{y=Y_t}}.
\end{equation}
\end{theorem}

\subsection{A special case}
We now apply the above to the market when the additional inside information at time $t$ is given by
\begin{equation}\label{eq4.4}
Y_t= (B(t+\varepsilon_t^{(1)}),B(t+\varepsilon_t^{(2)})); \quad t \in (0,T).
\end{equation}
for $\epsilon_t^{(1)}<\epsilon^{(2)}_t$.
We have, with $y=(y_1,y_2),$
\begin{align}\label{eq0.9}
&\mathbb{E}[\delta_Y (y)|\mathcal{F}_t]=(2\pi)^{-1}\exp(-\frac{(y_1-y_2)^2}{2(\epsilon_t^{(2)}-\epsilon_t^{(1)})})
\sqrt{\frac{1}{\epsilon_t^{(1)}(\epsilon_t^{(2)}-\epsilon_t^{(1)})}}\nonumber\\
&\exp(-\frac{1}{2\epsilon_t^{(1)}}(y_1-B(t))^2),
\end{align}
and
\begin{align}\label{eq0.10}
&\mathbb{E}[D_t\delta_Y (y)|\mathcal{F}_t]=(2\pi)^{-1}\exp(-\frac{(y_1-y_2)^2}{2(\epsilon_t^{(2)}-\epsilon_t^{(1)})})
\sqrt{\frac{1}{\epsilon_t^{(1)}(\epsilon_t^{(2)}-\epsilon_t^{(1)})}}\nonumber\\
&\exp(-\frac{(y_1-B(t))^2}{2\epsilon_t^{(1)}})\frac{y_1-B(t)}{\epsilon_t^{(1)}}.
\end{align}
Substituting \eqref{eq0.9} and \eqref{eq0.10} in\eqref{eq0.8}, we get
\begin{equation}\label{eq0.11}
\pi^*(t,y)=\frac{\alpha(t)}{\beta^2(t)}-\frac{B(t)-y_1}{\beta(t)\varepsilon_t^{(1)}}.
\end{equation}
Substituting \eqref{eq0.11} in \eqref{eq0.5} we get
\begin{align}\label{eq0.12}
&\int_{\mathbb{R}^2}\mathbb{E}[\int_0^T\{[\frac{\alpha(t)}{\beta^2(t)}-\frac{B(t)-y_1}{\beta(t)\varepsilon_t^{(1)}}]\beta(t)\mathbb{E}[D_{t^+}\delta_{Y_t}(y)|\mathcal{F}_t]+[\frac{\alpha(t)}{\beta^2(t)}-\frac{B(t)-y_1}{\beta(t)\varepsilon_t^{(1)}}]\alpha(t) \mathbb{E}[\delta_{Y_t}(y)|\mathcal{F}_t]\nonumber\\
&-\frac{1}{2}[\frac{\alpha(t)}{\beta^2(t)}-\frac{B(t)-y_1}{\beta(t)\varepsilon_t^{(1)}}]^2\beta^2(t)\mathbb{E}[\delta_{Y_t}(y)|\mathcal{F}_t]\}dt]dy\nonumber\\
&=\mathbb{E}[\int_0^T\{\int_{\mathbb{R}}(\int_{\mathbb{R}}(2\pi)^{-\frac{1}{2}}\sqrt{\frac{1}{(\epsilon_t^{(2)}-\epsilon_t^{(1)})}}
\exp(-\frac{(y_1-y_2)^2}{2(\epsilon_t^{(2)}-\epsilon_t^{(1)})})dy_2)\nonumber\\
&\times(2\pi\epsilon_t^{(1)})^{-\frac{1}{2}}
\exp(-\frac{(y_1-B(t))^2}{2\epsilon_t^{(1)}})[\frac{\alpha(t)}{\beta^2(t)}-\frac{B(t)-y_1}{\beta(t)\varepsilon_t^{(1)}}]\nonumber\\
&\Big(\beta(t)\frac{y_1-B(t)}{\epsilon_t^{(1)}}+\alpha(t)-\frac{1}{2}\beta^2(t)[\frac{\alpha(t)}{\beta^2(t)}-\frac{B(t)-y_1}{\beta(t)\varepsilon_t^{(1)}}]\Big)dy_1\}]\nonumber
\\
&=\frac{1}{2}(2\pi\varepsilon_t^{(1)})^{-\frac{1}{2}}\int_{\mathbb{R}}\mathbb{E}[\int_0^T\exp(-\frac{(B(t)-y_1)^2}{2\varepsilon_t^{(1)}})(\frac{B(t)-y_1}{\varepsilon_t^{(1)}}-\frac{\alpha(t)}{\beta(t)})^2 dt]dy_1.
\end{align}

Put
\begin{equation}
b=B(t) \text {  and  } \sigma=\frac{\alpha(t)}{\beta(t)}.
\end{equation}

Integrating the integrand above with respect to $dy_1$ we get:
\begin{align}
&(2\pi\varepsilon_t)^{-\frac{1}{2}}\int_{\mathbb{R}}\exp(-\frac{(B(t)-y_1)^2}{2\varepsilon_t^{(1)}})(\frac{B(t)-y_1}{\varepsilon_t^{(1)}}-\frac{\alpha(t)}{\beta(t)})^2dy_1\nonumber\\
&=(2\pi\varepsilon_t^{(1)})^{-\frac{1}{2}}\int_{\mathbb{R}}\exp(-\frac{(b-y_1)^2}{2\varepsilon_t^{(1)}})(\frac{b-y_1}{\varepsilon_t^{(1)}}-\sigma)^2dy_1  \nonumber\\
&=(2\pi\varepsilon_t^{(1)})^{-\frac{1}{2}}\int_{\mathbb{R}}\exp(-\frac{(b-y_1)^2}{2\varepsilon_t^{(1)}})(\frac{b-y_1}{\varepsilon_t^{(1)}}-\sigma)(\frac{b-y_1}{\varepsilon_t^{(1)}}-\sigma)dy_1\nonumber\\
&=(2\pi\varepsilon_t^{(1)})^{-\frac{1}{2}}\int_{\mathbb{R}}\{\frac{d}{dy_1}\exp(-\frac{(b-y_1)^2}{2\varepsilon_t^{(1)}})-\sigma\exp(-\frac{(b-y_1)^2}{2\varepsilon_t^{(1)}})\}(\frac{b-y_1}{\varepsilon_t^{(1)}}-\sigma)dy_1\nonumber\\
&=(2\pi\varepsilon_t^{(1)})^{-\frac{1}{2}}[\int_{\mathbb{R}}\frac{d}{dy_1}\exp(-\frac{(b-y_1)^2}{2\varepsilon_t^{(1)}})(\frac{b-y_1}{\varepsilon_t^{(1)}}-\sigma)dy_1+\sigma^2 \int_{\mathbb{R}}\exp(-\frac{(b-y_1)^2}{2\varepsilon_t^{(1)}}) dy_1]\nonumber\\
&=(2\pi\varepsilon_t^{(1)})^{-\frac{1}{2}}[\int_{\mathbb{R}}\exp(-\frac{(b-y_1)^2}{2\varepsilon_t^{(1)}})(\frac{1}{\varepsilon_t^{(1)}})dy_1+\sigma^2\int_{\mathbb{R}}\exp(-\frac{(b-y_1)^2}{2\varepsilon_t^{(1)}})dy_1]\nonumber\\
&=\frac{1}{\varepsilon_t^{(1)}}+\sigma^2=\frac{1}{\varepsilon_t^{(1)}}+(\frac{\alpha(t)}{\beta(t)})^2.
\end{align}

Then from equation \eqref{eq0.12} we get:
\begin{theorem}
The maximal expected logarithmic utility of Problem 1.1 in the insider market \eqref{eq1.3}, \eqref{eq4.4} with inside information $Y_t=(B(t+\epsilon^{1}_t),B(t+\epsilon^{2}_t))$ is
\begin{equation}
\mathbb{E}[\log X^{\pi^*}(T)]=\frac{1}{2}\mathbb{E}[\int_0^T(\frac{1}{\varepsilon_t^{(1)}}+(\frac{\alpha(t)}{\beta(t)})^2)dt].
\end{equation}
\end{theorem}
\begin{remark}
From this Theorem we get that the maximal expected utility of the terminal wealth depends only on the value of $B(t+\epsilon_t^{(1)})$. Thus the additional information about the value of $B(t+\epsilon^{(2)}_t)$ is irrelevant for the optimization problem if we already know $B(t+\epsilon^{(1)}_t).$ This is a strong indication that if the insider already knows the value of $B(t+\epsilon_t)$, then knowing in addition the values of $B(t+r)$ for all $r\in]\epsilon_t,\epsilon_0-t]$ does not increase the value of the optimal portfolio. However, we have not been able to prove this.
\end{remark}

As a corollary of Theorem 3.2, we get the following necessary condition for viability:

\begin{theorem}
This insider market \eqref{eq1.3}, \eqref{eq0.2}, \eqref{eq4.4} is viable only if
\begin{equation} \label{eq3.18}
\int_0^T\frac{1}{\varepsilon_t}dt<\infty.
\end{equation}
\end{theorem}

\begin{remark}
As noted in Remark 3.3, Theorem 3.2 supports our conjecture that we in fact have that the market is viable \emph{if and only if}  \eqref{eq3.18} holds.

\end{remark}

\begin{example}
\begin{enumerate}
  \item If $\epsilon_t=(T-t)^{q}$ for some $q>1$,  then
  \begin{equation}
  \int_0^T\frac{1}{\epsilon_t}dt=\int_0^T\frac{1}{(T-t)^q}dt=\infty.
  \end{equation}
  and the market is not viable. \\
  In particular, for $q=1$ (corresponding to $t+\epsilon_t=T$ for all $t$), this was first proved by \cite{PK} by different methods. In this case $B(\cdot)$ is a semimartingale with respect to $\mathbb{H}:=\mathbb{K}:=\{\mathcal{K}\}_{t \geq 0},$ where $ \mathcal{K}_t=\mathcal{F}_t \vee \sigma(B(T)), \forall t\in[0,T]$.
  \item If $\epsilon_t=(T-t)^p$ for some $p<1,$ then
  \begin{equation}
  \int_0^T\frac{1}{\epsilon_t}dt=\int_0^T(T-t)^{-p}dt=\frac{T^{1-p}}{1-p}<\infty.
  \end{equation}
 In view of Theorem 3.4 this indicates that the market is viable.
\end{enumerate}
\end{example}

\subsection{The case when $t+\epsilon_t \leq T$ for all $t$}

If $t+\epsilon_t \leq T$ for all $t$, then the natural corresponding information filtration for an insider with perfect memory is
\begin{equation}\label {eq3.21}
\mathbb{H}=\{\mathcal{H}_t\}_{t \geq 0} \text{ where } \mathcal{H}_t =\mathcal {F}_{t+\epsilon_t}.
\end{equation}
In this case $B(\cdot)$ is not a semimartingale with respect to $\mathbb{H}$ (see below).
Nevertheless, our calculation above shows the following:
\begin{theorem}
The insider market \eqref{eq1.3}, \eqref{eq3.21}, \eqref{eq4.4} is viable if and only if
 \begin{equation}
  \int_0^T\frac{1}{\epsilon_t}dt < \infty.
  \end{equation}
  Since in this case $\epsilon_t \leq T-t$ and $\int_0^T \frac{dt}{T-t} = \infty,$ we conclude that the market is never viable in this case.
  This is also a generalization of the \cite{PK} result, but in a different direction.
  \end{theorem}

\begin{proposition}
Suppose that $t\rightarrow \epsilon_t$ is of finite variation.
Then $B(\cdot)$ is not a semimartingale with respect to $\mathbb{H}$ where $\mathcal{H}_t=\mathcal{F}_{t+\epsilon_t}, \quad \forall t\geq0$.
\end{proposition}

\dproof
Suppose that $B(\cdot)$ is  an $\mathbb{H}$-semimartingale. Then
\begin{equation}
B(t)=\tilde{B}(t)+A(t),
\end{equation}
where $\tilde{B}(t)$ is an  $\mathbb{H}$- martingale and $A(t)$ is an $\mathbb{H}$-adapted  finite variation process.
Then, for $t+\varepsilon \leq t+h\leq T$,
\begin{align}
&0=\mathbb{E}[\tilde{B}(t+h)-\tilde{B}(t)|\mathcal{H}_t]\nonumber\\
&=\mathbb{E}[B(t+h)-B(t)|\mathcal{H}_t]-\mathbb{E}[A(t+h)-A(t)|\mathcal{H}_t]\nonumber\\
&=\mathbb{E}[B(t+h)-B(t)|\mathcal{F}_{t+\varepsilon_t}]-\mathbb{E}[A(t+h)-A(t)|\mathcal{H}_t]\nonumber\\
&=B(t+\varepsilon_t)-B(t)-\mathbb{E}[A(t+h)-A(t)|\mathcal{H}_t].
\end{align}
Then we get
\begin{equation}
B(t+\varepsilon_t)-B(t)=\mathbb{E}[A(t+h)-A(t)|\mathcal{H}_t].
\end{equation}
This is a contradiction, because $A$ is a finite variation process.

\fproof

\end{document}